\documentclass[journal,10pt]{IEEEtran}

\pdfoutput=1

\usepackage{amsmath}
\usepackage{amsfonts}
\usepackage{amssymb}

\usepackage{hyperref}

\usepackage{graphicx}
\usepackage{caption}
\usepackage{subcaption}
\usepackage{array}
\usepackage{xcomment}

\renewcommand{\eqref}[1]{Eq.~(\ref{#1})}
\newcommand{\figref}[1]{Fig.~\ref{#1}}
\newcommand{\secref}[1]{Section~\ref{#1}}
\newcommand{\tabref}[1]{Table~\ref{#1}}

\newcommand{\rmd}{\mathrm{d}}
\newcommand{\nind}[1]{{(#1)}}
\newcommand{\avrg}[1]{\langle#1\rangle}
\newcommand{\e}{\mathrm{e}}
\newcommand{\sync}{\text{sync}}
\newcommand{\Sync}{\text{SYNC}}

\newcommand{\new}[1]{#1}

\newcolumntype{x}[1]{>{\centering\let\newline\\\arraybackslash\hspace{0pt}}m{#1}}

\title{A guide to time-resolved and parameter-free measures of spike train synchrony}

\author{\IEEEauthorblockN{Mario Mulansky\IEEEauthorrefmark{1}, Nebojsa Bozanic\IEEEauthorrefmark{1}, Andreea Sburlea\IEEEauthorrefmark{2,3} and Thomas Kreuz\IEEEauthorrefmark{1}}\\[.5em]
\IEEEauthorblockA{\IEEEauthorrefmark{1}Institute for Complex Systems, CNR, Sesto Fiorentino, Italy}\\%
\IEEEauthorblockA{\IEEEauthorrefmark{2}BitBrain Technologies -- Research Department, Zaragoza, Spain}\\
\IEEEauthorblockA{\IEEEauthorrefmark{3}Institute of Investigation in Engineering of Aragon, I3A, University of Zaragoza, Spain}\\
\thanks{Email: \href{mailto:mario.mulansky@fi.isc.cnr.it}{\textsf{mario.mulansky@fi.isc.cnr.it}}}
\thanks{Email: \href{mailto:thomas.kreuz@cnr.it}{\textsf{thomas.kreuz@cnr.it}}}
}

\begin{document}

\maketitle

\begin{abstract}

Measures of spike train synchrony have proven a valuable tool in both experimental and computational neuroscience.
Particularly useful are time-resolved methods such as the ISI- and the SPIKE-distance, which have already been applied in various bivariate and multivariate contexts.
Recently, SPIKE-Synchronization was proposed as another time-resolved synchronization measure.
It is based on Event-Synchronization and has a very intuitive interpretation.
Here, we present a detailed analysis of the mathematical properties of these three synchronization measures.
For example, we were able to obtain analytic expressions for the expectation values of the ISI-distance and SPIKE-Synchronization for Poisson spike trains.
For the SPIKE-distance we present an empirical formula deduced from numerical evaluations.
These expectation values are crucial for interpreting the synchronization of spike trains measured in experiments or numerical simulations, as they represent the point of reference for fully randomized spike trains.

\end{abstract}


\section{Introduction}
Understanding the details of information processing in the brain is one of the most challenging and exciting problems of our time. 
It has been widely established that the brain can be considered as an enormous network of spiking neurons, and that the spikes convey the information processed within this network~\cite{kandel2012principles, quiroga2013principles}.
In the first studies, it was assumed that the information is encoded in the spike rates of the neurons, and several experiments confirmed this assumption~\cite{perkel1968coding}.
However, it became clear rather soon that in many, typically more complex situations the rate coding is not sufficient to represent the available information.
For example, it was found that even single spike events can be responsible for the discrimination between different stimuli~\cite{rieke1996spikes}.
Hence, solely studying the spike rates is not sufficient for unraveling the neural code -- exact spike timings need to be analyzed.

In the last two decades various spike train distances, some inspired by existing mathematical distance measures, have been proposed~\cite{victor1996nature, rossum2001novel, quiroga2002event, schreiber2003new, schrauwen2007linking, kreuz2007measuring, kreuz2009measuring, naud2011improved, kreuz2011time, kreuz2013monitoring, rusu2014new, muino2014frequent}.
Their application to real neural data has led to a remarkable increase in the understanding of neural networks and neural coding~\cite{victor2005spike}.

An important step was the introduction of time-resolved synchrony measures that allow to analyze the time dependence of spike train similarities~\cite{quiroga2002event, kreuz2007measuring, kreuz2013monitoring}.
With this ability, it is now possible to investigate synchrony changes of pairs or groups of neurons, for example induced by external or internal stimuli.
Hence, such time-resolved synchronization profiles open new opportunities in spike train analysis.

However, before proceeding with a detailed analysis of spike train similarity, first one has to obtain the spike events from the experimental measurements.
This process is typically called spike detection and its difficulty is often underrated.
Therefore, we give a brief overview of the existing techniques of spike and event detection in \secref{sec:detection}, which represents the basis for the analysis of spike train synchrony discussed in \secref{sec:distance}.
There, we present three time-resolved synchrony measures: the ISI-distance~\cite{kreuz2007measuring}, the SPIKE-distance~\cite{kreuz2013monitoring} as well as a new, time-resolved variant of Event-Synchronization~\cite{quiroga2002event}, called SPIKE-Synchronization~\cite{kreuz2015SPIKY}.
The ISI- and SPIKE-distance, discussed several times in the past and being well formalized, still miss a comprehensive analysis of their mathematical properties.
Here, we will fill this gap and provide mathematical details of the ISI-distance in \secref{sec:isi_distance} and the SPIKE-distance in \secref{sec:spike_distance}.
\secref{sec:spike_sync} introduces SPIKE-Synchronization, including a detailed analysis of its mathematical properties.
Finally, \secref{sec:summary} contains a brief summary and our conclusions.

\section{Detection Methods} \label{sec:detection}
Experiments on brain activity, whether \textit{in vivo} or \textit{in vitro}, are  typically done by placing electrodes in the region of interest.
Most common techniques provide intracellular recordings of a membrane potential of a single isolated neuron (single-unit recordings) or measure the mean extracellular field potential generated by electrical activities of several (nearby) neurons (multi-unit recordings).
The fundamental observables are the spikes, also called action potentials, emitted by the neurons and propagated to target nodes of the neural network via synapses.
As the name suggests, these spike events are rapid, distinct maxima of the membrane potential.
The prerequisite for any spike train analysis is the extraction of the spike times from the measured membrane potential.
The set of spike times of a single neuron is then called the \emph{spike train}, and is defined as:
\begin{equation}
 s = \{t_i\}, \quad\text{with}\quad t_i<t_{i+1},
\end{equation} 
where the $t_i$ represent the times of the spikes.
Below we give a brief introduction of the current spike and event detection methods before proceeding to the measures of spike train synchrony.

\subsection{Spike Detection and Sorting}

\begin{figure}
\centering
 \includegraphics[width=0.49\textwidth]{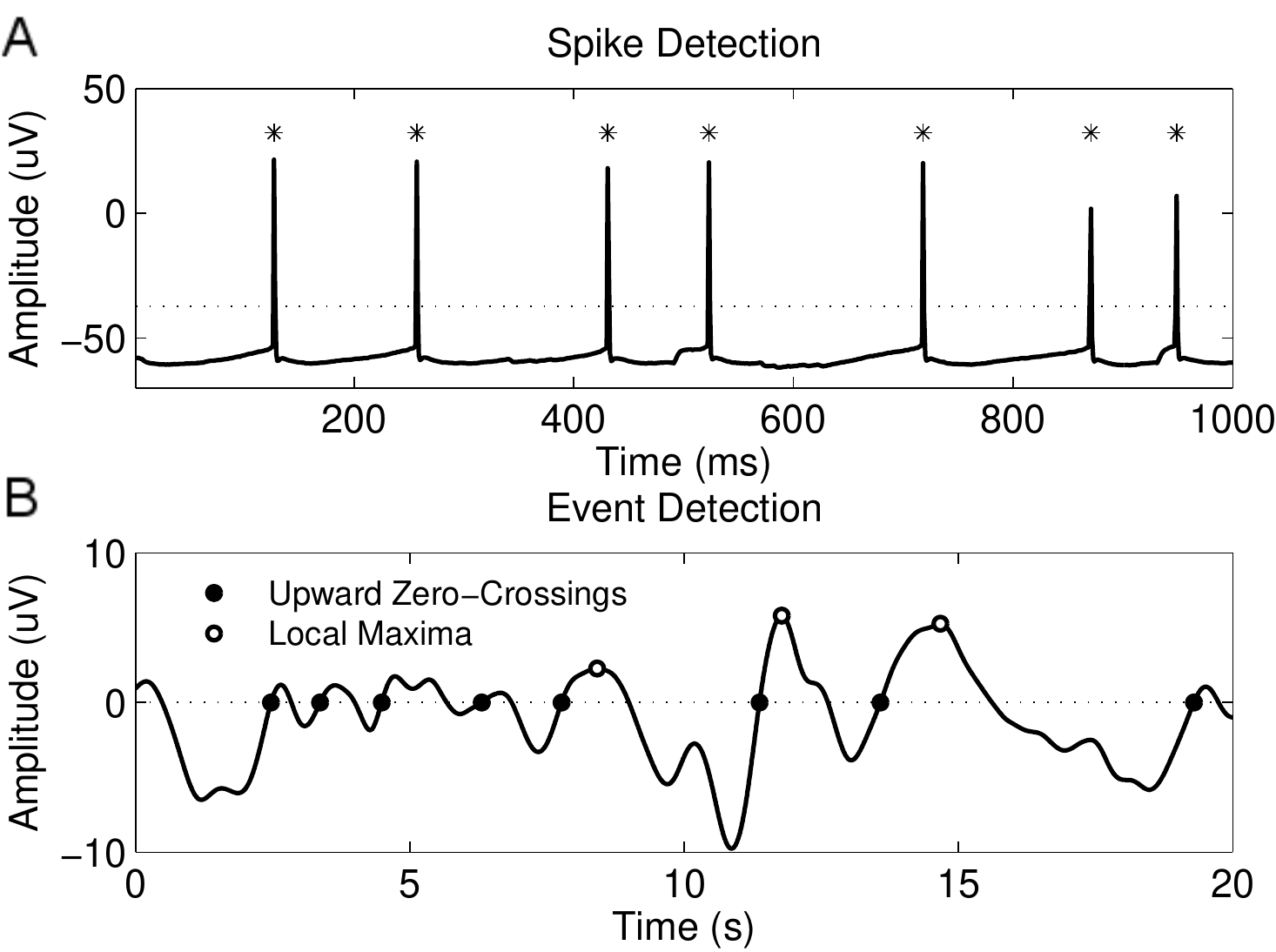}
\caption{Panel A shows a spike detection scenario in which the dotted line is an adaptive threshold equal to three standard deviations of the signal (neuronal membrane potential). Panel B  shows the detection of two different types of events -- \new{upward} zero-crossings and local maxima above one standard deviation of the signal (EEG).
\label{fig:detection}}
\end{figure} 

Spike detection means to detect discrete spiking events in continuous profiles of membrane or extracellular field potential.
Most commonly, the first step is to increase the signal to noise ratio by enhancing the spike waveform and reducing the noise.
The simplest way of determining the spike locations is a threshold detector~\cite{mcdonough1995detection} (cf.\ \figref{fig:detection}A).
Many spike detection techniques  are based on an amplitude threshold with little or no preprocessing and either static or adaptive thresholds~\cite{gozani1994optimal, guillory1999100, chandra1997detection, lewicki1998review}.
In many experimental situations where the extra-cellular field potential is measured, the profiles contain spike events from more than one neuron.
Therefore, one not only has to identify the spikes, but also assign them into groups corresponding to the different neurons.

In cases of intracellular recording or isolated cells however, one can assume that all spikes of one recording belong to a single neuron.
There exist several techniques of spike detection for this situation, the most important ones are briefly described below.

\emph{Static threshold} detectors rely either on a single threshold for the detection of one edge~\cite{gozani1994optimal} or on several thresholds to detect the slopes of the spike~\cite{guillory1999100} or other characteristics such as skewness, width, area under the peak, etc.
Although simple thresholding is attractive for real-time implementations due to its computational simplicity, it is thought to be too sensitive to noise and often requires user input to set effective threshold levels~\cite{lewicki1998review}.
Overlapping spikes further reduce the efficacy of simple threshold detectors~\cite{obeid2004evaluation}.

\emph{Adaptive thresholds} are dependent on the nature of the signal and are particularly useful in multiple channel recordings, allowing optimization in several channels simultaneously. 
The threshold levels are often based on the signal's standard deviation~\cite{pouzat2002using} or the standard deviation of the noise computed with respect to the median of the signal in which the spiking activity has the smallest contribution~\cite{quiroga2004unsupervised}.
 
\emph{Energy-based} spike detectors first apply a nonlinear operation to the signal before using static or adaptive threshold detection~\cite{kim2000neural, obeid2004evaluation}.
This nonlinear energy operator estimates the square of the instantaneous product of amplitude and frequency of a sufficiently sampled signal~\cite{kaiser1990simple} and increases the separation of the spikes from the background noise.

Often, one faces the more complicated situation where the measured signal possibly contains spike events from several neurons (multi-unit recordings).
Then, additionally to spike detection also spike sorting must be performed.
This involves isolating the neural signals and assigning each recorded waveform to the neuron of origin~\cite{lewicki1998review}.
Spike sorting techniques are based on the fact that action potentials recorded from the same cell tend to have a stereotypical spike shape determined by the cell's morphology and biophysical properties, but also by its position relative to the recording electrode~\cite{quiroga2007spike}.
After the spike events have been detected using the techniques described above, spike sorting is usually done in two steps.

The first step is feature extraction, where a number of significant features are obtained from the spike events, which allow to separate the different clusters afterwards.
The most prominent method used for feature extraction is \emph{wavelet analysis}~\cite{quiroga2004unsupervised,hulata2002method}, which was shown to outperform the previously used \emph{principal component analysis (PCA)}~\cite{pavlov2007sorting}.
 
The second step is clustering, where the spikes are assigned into different groups corresponding to different neurons.
The clustering is based on the features obtained before.
Typical approaches include \emph{expectation maximization}~\cite{harris2000accuracy}, \emph{super-paramagnetic clustering (SPC)}~\cite{quiroga2004unsupervised} or \emph{support vector machines (SVN)}~\cite{vogelstein2004spike}.

It should be noted that despite the progress made in recent years, spike sorting remains a hard problem.
Typical issues are the drifting of neurons, overlapping spikes and neurons with very similar spike waveforms.
Increasing the number of observation electrodes or simultaneous intra- and extra-cellular recordings improve the results, but even then for experimental data it is usually impossible to verify if the spikes were assigned correctly.

\subsection{Event Detection}

Spike detection is the most prominent example for a transformation from continuous data to a discrete point process. 
This transformation becomes more complicated if the continuous time series do not contain any pronounced and stereotypical events such as spikes. 
In this case it is possible to use rather generic events such as zero crossings, local maxima and minima or any other particular feature characteristic of the signal~\cite{quiroga2002event}. 
An example of an electroencephalogram (EEG) trace with two different types of events is shown in \figref{fig:detection}B.

\renewcommand{\arraystretch}{1.5} 
\begin{table*}[t]
 \begin{tabular}{m{0.15\textwidth} | x{0.25\textwidth} | x{0.25\textwidth} | x{0.25\textwidth} }
  & {\normalsize ISI: $I(t)$} & {\normalsize SPIKE: $S(t)$} & {\normalsize SPIKE-Sync: $C_k$}\\
  \hline
  Type of profile & piecewise constant function & piecewise linear function & discrete function \\
  \hline
  Relation between profile and overall measure & $D_I = \frac1T \int I(t)\rmd t$ & $D_S = \frac1T\int S(t)\rmd t$ & $\text{SYNC} = \frac1M \sum_k C_k$,\newline[.5em] $D_\sync = 1-\text{SYNC}$\\
  \hline
   
  Range & $0\leq I(t),D_I \leq1$ & $0\leq S(t),D_S \leq1$ & $0\leq C_k, \text{SYNC},D_\sync \leq1$\\
  \hline
  
  Value for identical spike trains $s_1=s_2$ & $I(t)=0$, $D_I=0$ & $S(t)=0$, $D_S=0$ & $C_k$, $\text{SYNC}=1$, $D_\sync = 0$.\\
  \hline
  
  
  Expectation value for Poisson spike trains with $r=\lambda_1/\lambda_2$ & $\avrg{D_I} = \frac{1}{(1+r)^2} + \frac{1}{(1+r^{-1})^2}$ & $\avrg{D_S}\approx \frac12 -\frac15\e^{-(\log r)^2/8}$ & $\avrg{\text{SYNC}} = \frac1{r^{-1}+r+2}$,\newline[.5em] $\avrg{D_\sync} = 1 - \avrg{\text{SYNC}}$
 \end{tabular} 
\caption{Overview of the properties of different spike train synchronization measures. \label{tab:distance_properties}}
\end{table*}

The appropriate definition of the event is crucial since the information from the continuous time series is reduced to the time stamps of the events. 
It is not the synchronization between the time series as a whole that is evaluated, but rather the synchronization between the defined events only. 
Different choices of events can yield different results.

\section{Distance Measures} \label{sec:distance}

To quantify the degree of synchronization of two spike trains $s_1$ and $s_2$, a distance measure~$D$ is introduced to map the pair of spike trains into a positive number representing the differences between these spike trains.
The normalization of $D$ is arbitrary, but a sensible choice is to limit the distance to the interval $[0,1]$, i.e.:
\begin{equation}
 D:\{s_1,s_2\} \mapsto [0,1],
\end{equation} 
where $D=0$ represents identical spike trains ${s_1=s_2}$, while larger values denote a higher degree of dissimilarity.
Many distance measures rely on one (or more) parameters that have to be chosen appropriately for the given spike trains and are usually connected to the typical time-scale of the spike events, for example the Victor-Purpura and van Rossum distances~\cite{victor1996nature, rossum2001novel}.
Clearly, the existence of such a parameter introduces ambiguity in the analysis as its optimal value is a priori unknown, furthermore it is even unclear how to define such an ``optimal'' parameter value~\cite{chicharro2011what}.

To overcome this problem, \emph{parameter-free} distance measures have been introduced, namely the ISI- and the SPIKE-distance as well as SPIKE-Synchronization described below.
The ISI- and SPIKE-distance are based on time dependent distance profiles $I(t)$, $S(t)$, from which the overall spike train distance can be computed via simple integration, e.g.:
\begin{equation} \label{eqn:int_profile}
 D_S = \frac1{T}\int_{0}^{T} S(t) \rmd t.
\end{equation} 
The time $T$ denotes the duration of the spike trains, i.e.\ the recording interval.
Having such a profile $S(t)$ allows for a time resolved analysis of the spike train synchrony, which is important for example to detect synchronization triggered by external or internal events.

So far, we have only introduced a bivariate distance, that is the distance between two spike trains $s_1$ and $s_2$.
But usually one has to deal with multiple spike trains, e.g.\ from simultaneous measurements of several neurons.
Bivariate distance can be extended to multivariate distance by simply averaging the bivariate distances of all pairs of spike trains.
Suppose we have $N$ spike trains, then the averaged multivariate distance is computed as:
\begin{equation}
 D^a = \frac2{N(N-1)} \sum_{n=1}^{N-1}\sum_{m=n+1}^{N} D_{n,m},
\end{equation} 
where $D_{n,m} = D(s_n,s_m)$ is the distance between the spike trains $s_n$ and $s_m$.
As this average commutes with the time integration, we can also readily introduce a multivariate distance profile:
\begin{equation} \label{eqn:avrg_profile}
 S^a(t) = \frac2{N(N-1)} \sum_{n=1}^{N-1}\sum_{m=n+1}^{N} S_{n,m}(t),
\end{equation} 
where still $D^a = \int S^a(t) \rmd t/T$.

This completes the framework of time-resolved spike train distances.
In the following we will provide a detailed description of the ISI- and SPIKE-distance and discuss their properties.
Additionally, the recently developed SPIKE-Synchronization is presented, another time-resolved measure based on the well-known Event-Synchronization~\cite{quiroga2002event}.
Despite also being time-resolved, SPIKE-Synchronization has some fundamental mathematical differences compared to the ISI- and SPIKE-distance, as will be explained later.
An overview of the mathematical properties of all three measures is shown in \tabref{tab:distance_properties}.

\subsection{ISI-Distance} \label{sec:isi_distance}

The ISI-dissimilarity-profile~$I(t)$, introduced by Kreuz et al.~\cite{kreuz2007measuring}, was the first parameter-free, instantaneous measure of spike train synchrony.
It relies on the ratio between the concurrent interspike intervals of the two spike trains.

\subsubsection{Definition} 
Let $\{t^{(1)}_i\}$ be the spike times of the first spike train, then the interspike intervals are defined as:
\begin{equation} \label{eqn:interspike_intervals_index}
 \nu^{(1)}_i = t^{(1)}_{i+1}-t^{(1)}_{i},
\end{equation}
and similarly $\nu^{(2)}_i$ for the second spike train.
To arrive at a time dependent profile, the sequences of interspike intervals are transformed into piecewise constant functions:
\begin{equation} \label{eqn:interspike_intervals}
 \nu^{(1),(2)}(t) = \nu^{(1),(2)}_i\quad\text{for}\quad t^{(1),(2)}_i\leq t < t^{(1),(2)}_{i+1},
\end{equation}
which for each interval $[t_{i},t_{i+1})$ takes the value $\nu_i$.
\figref{fig:spike_definitions} shows an example of two spike trains with the definition of $\nu^{\nind{1},\nind{2}}(t)$ at some time $t$.
The ISI-profile is then given as the normalized absolute difference of the interspike intervals:
\begin{equation} \label{eqn:isi_profile}
 I(t) = \frac{|\nu^{(1)}(t) - \nu^{(2)}(t)|}{\max\{\nu^{(1)}(t),\nu^{(2)}(t)\}},\quad t \in [0,T].
\end{equation}
The interval $[0,T]$ again denotes the observation period of the two spike trains, i.e.\ $0\leq t^{(1),(2)}_i \leq T$.
However, there is a potential ambiguity concerning the first and the last interspike interval if the first (last) spikes do not coincide with the start (end) time of the observation interval, e.g.\ $t^{(1),(2)}_1 \neq 0$.
\new{As an improvement to previous definitions~\cite{kreuz2007measuring}, we here introduce an edge-correction for the ISI-distance to estimate the first and last interspike interval~\cite{kreuz2015SPIKY}.
Therefore, for the first interspike interval we use the maximum of the distance between the start of the observation interval and the first spike, and the first known interspike interval: ${\nu(t<t_1) = \max\{t_1, t_2-t_1\}}$.
A similar estimation is performed for the very last interspike interval. 
This makes the ISI-profile~\eqref{eqn:isi_profile} well defined for the whole observation interval $t\in[0,T]$.
}
As the interspike intervals are piecewise constant functions, also the ISI-profile is piecewise constant.
The ISI-distance is then given as the integral over the whole profile, cf.\ \eqref{eqn:int_profile}.

\subsubsection{Properties}
From the normalization in \eqref{eqn:isi_profile}, it is immediately obvious that the ISI-profile is bounded by ${0\leq I(t)<1}$, and the value $I(t)=0$ is only obtained for identical interspike intervals ${\nu^{(1)}(t) = \nu^{(2)}(t)}$.
Consequently, also the ISI-distance is bounded: ${0\leq D_I<1}$.
The ISI-distance is clearly symmetric: $D_I(s_1,s_2) = D_I(s_2,s_1)$ by construction.
Furthermore, it can be shown that the ISI-distance also fulfills the triangle inequality: $D_I(s_1,s_2) + D_I(s_2,s_3) \geq D_I(s_1,s_3)$. A proof is sketched in Appendix~B of~\cite{lyttle2011new}.
\new{For identical spike trains it is obvious that the ISI-distance vanishes: $D_I=0$.
However, for two spike trains with constant and equal interspike intervals $\nu^{(1)}(t) = \nu^{(2)}(t) = \nu$, but with a global shift, the ISI-distance also evaluates to zero.
Due to this degeneracy, the ISI-distance is only a \emph{pseudo-metric}, but a full metric space can be recaptured  by considering all degenerate spike trains with the same constant interspike interval, but overall time shifts as an equivalence class.
}

\figref{fig:profiles} shows exemplarily a multivariate ISI-profile of ${N=50}$ spike trains.
The change from the noise dominated spikes to increasingly synchronous events is captured quite well by the ISI-profile.
In the beginning, we observe ISI-values very close to those expected for random Poisson spike trains, while the values drop significantly in the second half which indicates higher similarity.
However, the ISI-profile is unable to detect the additional synchronous events within the random spike trains at the beginning.
This is due to the fact that the ISI-profile only incorporates information about interspike intervals, and not about exact spike timings.

For the ISI-distance it is possible to compute the expectation value $\langle D_I\rangle$ for two Poisson spike trains analytically.
Note, that this expectation value depends only on the ratio of the rates of the two spike trains: $r=\lambda_1/\lambda_2$.
A straightforward calculation of this value is performed in Appendix~\ref{app:isi_dist_poisson} and gives:
\begin{equation} \label{eqn:D_isi_poisson}
 \langle D_I(r)\rangle = \frac{1}{(1+r)^2} + \frac{1}{(1+r^{-1})^2}.
\end{equation} 
For two Poisson spike trains with equal rates, $r=1$, we thus find $\langle D_I\rangle = 1/2$, while in the limit where one spike train is much faster than the other one, $r\rightarrow0,\infty$ we have $\langle D_I\rangle\rightarrow1$.
This is visualized in \figref{fig:poisson_distance}.

\subsection{SPIKE-Distance} \label{sec:spike_distance}
The SPIKE-dissimilarity-profile~$S(t)$, first introduced in~\cite{kreuz2011time} and subsequently improved in~\cite{kreuz2013monitoring}, provides a time-resolved distance measure that relies on the exact timings of spike events.

\subsubsection{Definition}
\begin{figure}
\centering
 \includegraphics[width=0.49\textwidth]{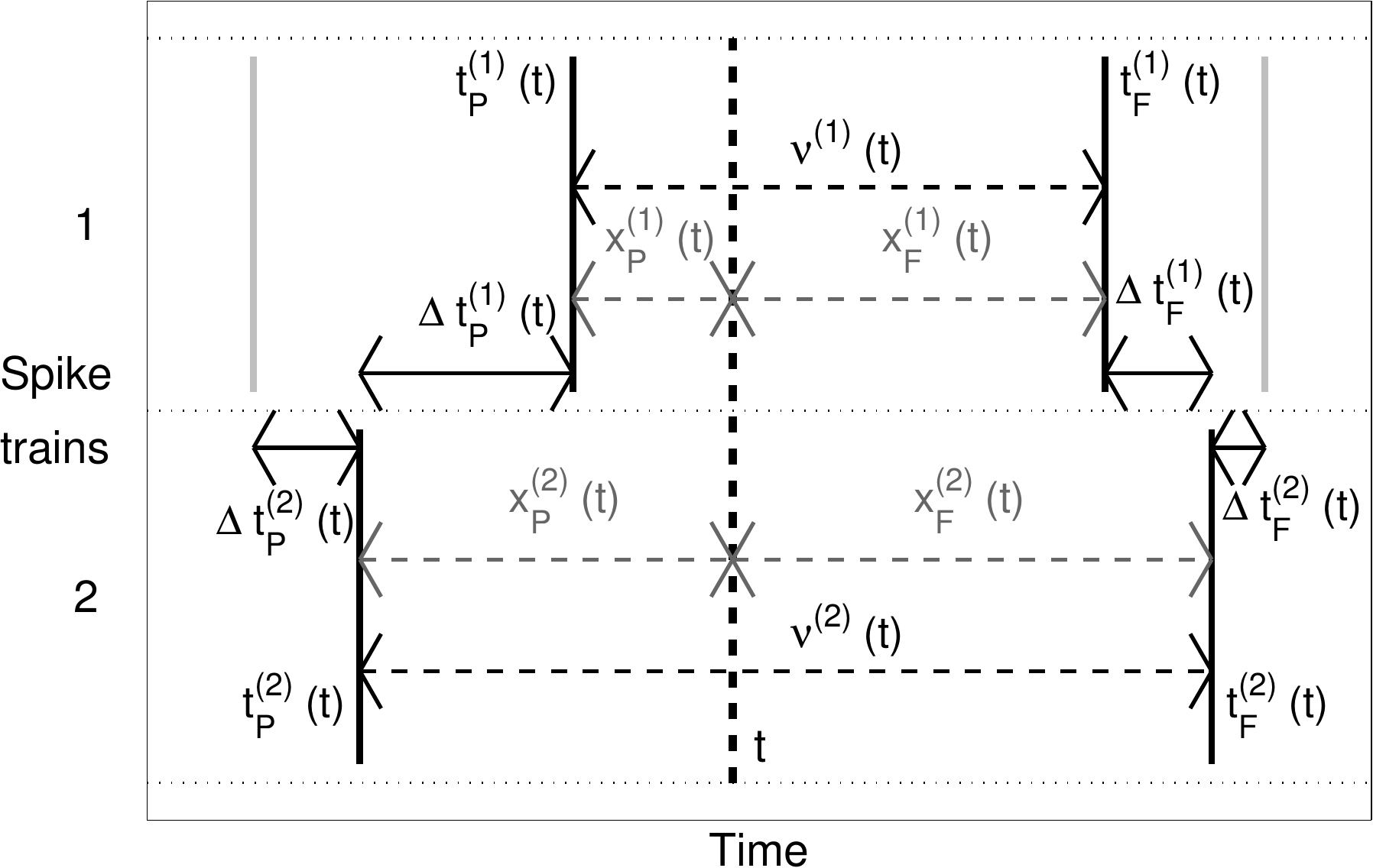}
\caption{Local definitions of interspike intervals and time differences required for the calculation of the ISI- and the SPIKE-profile.
\label{fig:spike_definitions}}
\vspace{-0.2cm}
\end{figure}

The computation of $S(t)$ is based on the four corner spikes surrounding the current time $t$: the preceding spikes $t_P^{(1),(2)}(t)$ and the following spikes $t_F^{(1),(2)}(t)$ of each spike train (cf.\ \figref{fig:spike_definitions}).
The current interspike intervals can be expressed in terms of these corner spikes as ${\nu^{\nind{1},\nind{2}}(t)=t_F^{(1),(2)}(t)-t_P^{(1),(2)}(t)}$.
For each of the corner spikes, the distance to the closest spike of the \emph{other spike train} is computed, e.g.:
\begin{equation}
 \Delta t_P^\nind{1}(t) = \min_i\{|t_P^\nind{1}-t^\nind{2}_i|\},
\end{equation} 
and similarly for $\Delta t_P^\nind{2}$ and $\Delta t_F^{\nind{1},\nind{2}}$.
Note, that by definition $\Delta t_{P,F}^{\nind{1},\nind{2}}(t)$ are piecewise constant functions.
\figref{fig:spike_definitions} shows a graphical representation of these definitions.
These distances are then weighted by the distance of the corner spikes to the current time by the weighting factors:
\begin{equation}
 x_P^{\nind{1},\nind{2}}(t) = t-t_P^{\nind{1},\nind{2}} \quad\text{and}\quad x_F^{\nind{1},\nind{2}}(t) = t_F^{\nind{1},\nind{2}}-t,
\end{equation}
with $x_P^{\nind{1},\nind{2}}(t) + x_F^{\nind{1},\nind{2}}(t) = \nu^{\nind{1},\nind{2}}(t)$, as seen from \figref{fig:spike_definitions}.
The weighted distance for the first spike train then reads:
\begin{equation} \label{eqn:spike_weighted_dist}
 S_1(t) = \frac{\Delta t_P^\nind{1}(t) x_F^\nind{1}(t) + \Delta t_F^\nind{1}(t) x_P^\nind{1}}{\nu^\nind{1}(t)},
\end{equation} 
and similarly $S_2(t)$ is defined for the second spike train.
As $x_{P,F}^{\nind{1},\nind{2}}(t)$ are linear in $t$, the resulting functions $S_{1,2}(t)$ are piecewise linear, with possible jumps at the interval edges $t^{\nind{1},\nind{2}}_i$.
Finally, these local distances are then weighted by the local interspike intervals and with a proper normalization we arrive at the definition of the SPIKE-profile:
\begin{equation} \label{eqn:spike_profile}
 S(t) = \frac{S_1(t) \nu^\nind{2}(t) + S_2(t) \nu^\nind{1}(t)}{\frac12(\nu^\nind{1}(t)+\nu^\nind{2}(t))^2}, \quad t \in [0,T],
\end{equation}
which again is a piecewise linear function as $S_{1,2}$ are piecewise linear while the other terms are piecewise constant.

As for the ISI-profile above, we face a potential ambiguity for the first and the last interval.
\new{Previous definitions of the SPIKE-distance introduced auxiliary spikes at the edges, which lead to spurious synchrony.
An improved treatment of the edges, similar to the edge correction for the ISI-distance above, is presented in \cite{kreuz2015SPIKY}.
}

\subsubsection{Properties}
The normalization in \eqref{eqn:spike_profile} again ensures that the SPIKE-profile is bound to ${0 \leq S(t) < 1}$.
Hence, the same bounds hold for the SPIKE-distance ${0\leq D_S = \int S(t)\rmd t<1}$.
Furthermore, $D_S(s_1,s_2)=0$ only if $s_1=s_2$, and the SPIKE-distance is also symmetric $D_S(s_1,s_2)=D_S(s_2,s_1)$.
However, it is not a metric as it is possible to construct spike trains that violate the triangular inequality: $D_S(s_1,s_2) + D_S(s_2,s_3) \ngeq D_S(s_1,s_3)$.

\begin{figure}[t]
  \centering
    \includegraphics[width=0.48\textwidth]{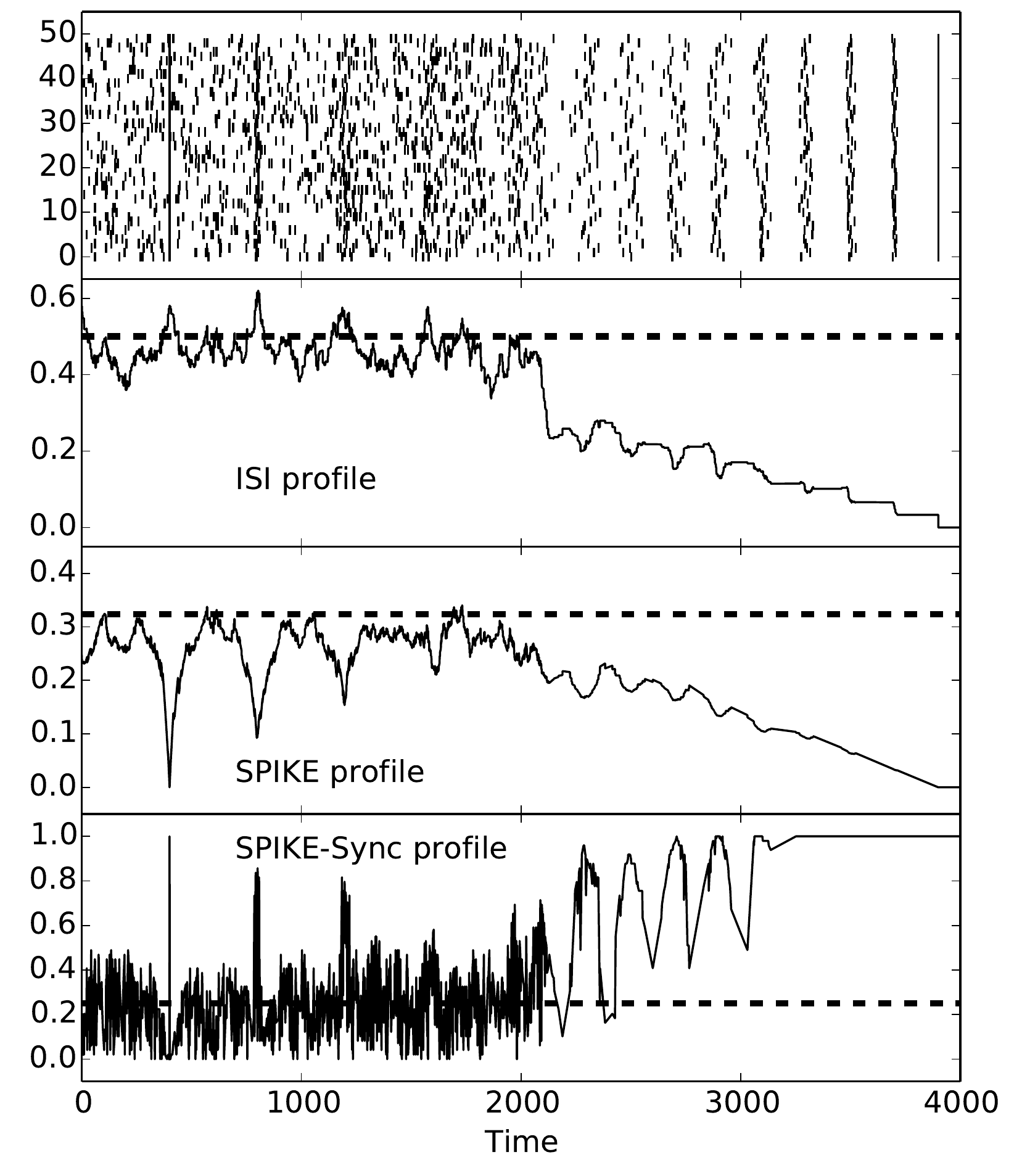}
  \caption{Multivariate ISI, SPIKE and SPIKE-Synchronization profiles for $M=50$ spike trains (shown in top panel). The spike trains are generated artificially such that the first half consists of a noisy background with a few synchronous events with increasing jitter, while the second half contains increasingly synchronized events.
  The dashed lines represent the respective expectation values for Poisson spike trains.}
   \label{fig:profiles}
\end{figure}

\figref{fig:profiles} shows an exemplary SPIKE-profile for the same $N=50$ spike trains as before.
As for the ISI-profile, the overall behavior of noise dominated spike times in the beginning and the increasing synchronization afterwards is well captured by the SPIKE-profile.
Additionally, the SPIKE-distance, relying on exact spike timings, is able to detect the synchronization events in the beginning as seen from the clear minima in the SPIKE-profile.

Unfortunately, the complicated definition of the SPIKE-distance makes an analytic computation of the expectation value $\avrg{D_S}$ for two Poisson spike trains intractable.
However, it is clear that also the SPIKE-distance should only depend on the rate ratio $r=\lambda_1/\lambda_2$ of the two Poisson spike trains and approach the value $D_S=1/2$ in the limit $r\rightarrow0,\infty$.
This is seen in \figref{fig:poisson_distance}.
Similarly to the ISI-distance and SPIKE-Synchronization below, the SPIKE-distance exhibits a clear minimum for spike trains with equal rates $\lambda_1=\lambda_2$, i.e.\ $r=1$.
As we are unable to obtain an exact analytic result for $\avrg{D_S}$ at this point, we provide an empirical approximation.
Therefore, we use the following function that already incorporates the properties mentioned above ($\lim_{r\rightarrow0,\infty} \avrg{D_S}=0.5$, minimum at $r=1$):
\begin{equation} \label{eqn:D_spike_avrg}
 \avrg{D_S} = \frac12 -\alpha\e^{-(\log r)^2/(2\beta^2)}.
\end{equation} 
From visual inspection, we find that with $\alpha=0.2$, $\beta=2$, \eqref{eqn:D_spike_avrg} provides an excellent approximation of the average SPIKE-distance for two Poisson spike trains as shown in \figref{fig:poisson_distance}.

\subsection{SPIKE-Synchronization} \label{sec:spike_sync}
SPIKE-Synchronization can be understood as an instantaneous coincidence detector.
It was recently proposed by Kreuz et al.~\cite{kreuz2015SPIKY} and is derived from Event-Synchronization~\cite{quiroga2002event}.
In contrast to the ISI- and SPIKE-distance, SPIKE-Synchronization quantifies similarity instead of difference, but a distance measure can be constructed in a straightforward way as shown below.

\subsubsection{Definition}
For the SPIKE-Synchronization profile, a coincidence indicator~$C^{\nind{1},\nind{2}}_i$ is defined for every spike of the two spike trains $s^{\nind{1},\nind{2}}$.
This coincidence indicator can have two possible values: $C_i=1$ if the spike at $t_i$ is part of a coincidence, and $C_i=0$ if not.
Similar as for Event-Synchronization, this coincidence indicator is given by:
\begin{equation} \label{eqn:spike_sync_cn}
 C^\nind{1}_i = \begin{cases} 1\quad\text{if}\quad \min_j(|t^\nind{1}_i - t^\nind{2}_j|) < \tau^{\nind{1,2}}_{ij}\\
        0\quad\text{otherwise}.
       \end{cases}
\end{equation} 
That means, a coincidence is found for the spike $i$ of the first spike train if the distance to the closest spike $j$ of the second spike train is smaller than the coincidence window $\tau^{\nind{1,2}}_{i,j}$.
The coincidence window is defined adaptively according to the local firing rate:
\begin{equation} \label{eqn:spike_sync_tau}
 \tau^{\nind{1,2}}_{ij} = \frac12\min\{\nu^\nind{1}_{i}, \nu^\nind{1}_{i-1}, \nu^\nind{2}_{j}, \nu^\nind{2}_{j-1}\},
\end{equation} 
with $\nu^{\nind{1},\nind{2}}$ being the interspike intervals as given in \eqref{eqn:interspike_intervals_index}.
The coincidence indicator for the second spike train $C^\nind{2}_i$ is computed in the same way as in \eqref{eqn:spike_sync_cn} but with exchanged indices $\nind{1}\leftrightarrow\nind{2}$.

The SPIKE-Synchronization profile is then obtained by merging the coincidence indices of the two spike trains ${\{C_k\} = \{C^\nind{1}_i\}\cup\{C^\nind{2}_i\}}$ as well as the spike times: ${\{t'_k\} = \{t^\nind{1}_i\}\cup\{t^\nind{2}_i\}}$, which results in a discrete function defining a SPIKE-Synchronization value for each spike time of the two spike trains: $t'_k \mapsto C_k$.
For visual purposes, it might be reasonable to connect the individual points of this discrete profile, but opposed to the ISI- or SPIKE-profile above, SPIKE-Synchronization is only defined exactly for the spike times -- any connecting lines do not represent real intermediate values but are merely a guide to the eye.
Note that $C_k=1$ for all $k$ means that every spike is part of a coincidence, while $C_k=0$ means no synchronous spikes appeared.

\begin{figure}
\centering
 \includegraphics[width=0.48\textwidth]{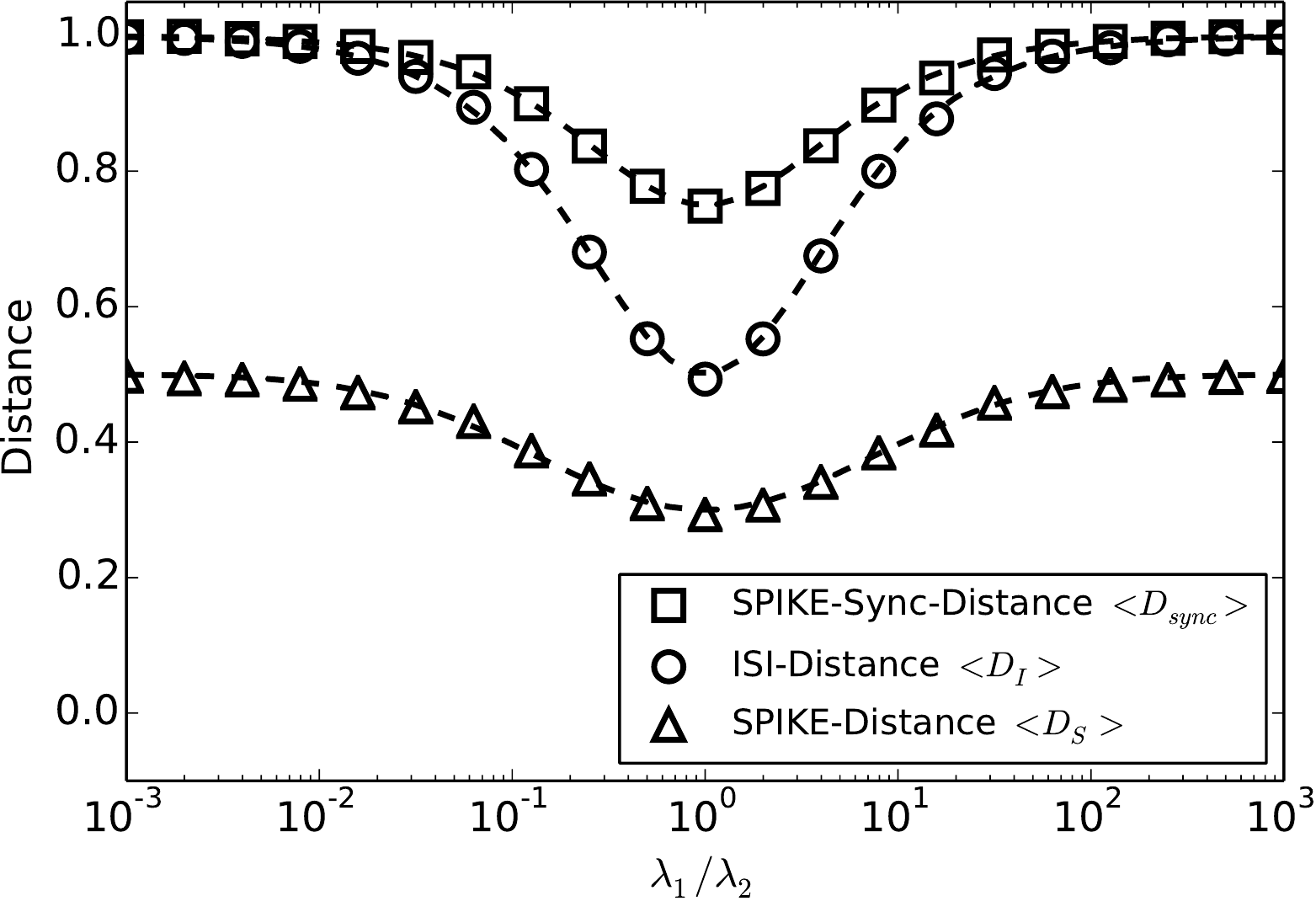}
\caption{Numerical results for the ISI-distance (circles), SPIKE-distance (triangles) and SPIKE-Synchronization-distance (squares) of two Poisson spike trains with a total of $M\approx20000$ spikes in dependence of the rate ratio $r=\lambda_1/\lambda_2$. The black dashed lines represent the analytical result for the ISI-distance \eqref{eqn:D_isi_poisson}, the empirical curve for the SPIKE-distance \eqref{eqn:D_spike_avrg} and the analytical result for the SPIKE-Synchronization distance~\eqref{eqn:D_sync_avrg}.
\label{fig:poisson_distance}}
\end{figure}

Being a discrete function, integrals are not defined for the SPIKE-Synchronization profile.
However, an overall SPIKE-Synchronization value~$\text{SYNC}$ can be obtained very naturally by summation:
\begin{equation} \label{eqn:overall_spike_sync}
 \text{SYNC} = \frac1M \sum_{k=1}^M C_k = \frac C M,
\end{equation}
where $M$ is the total number of spikes in the merged spike train~$\{t'_k\}$ and $C$ denotes the total number of coincident spikes.
As seen from \eqref{eqn:overall_spike_sync}, this value has a very intuitive interpretation as it simply represents the fraction of coinciding spikes.
The coincidence factor introduced in~\cite{kistler1997reduction} and further analyzed in~\cite{naud2011improved} has a similar interpretation.
However, it is not time resolved and its coincidences are defined with a constant coincidence window $\tau=\text{const}$.
Note, that $\text{SYNC}$ quantifies similarity, but a distance measure can be trivially obtained:
\begin{equation}
 D_\text{sync} = 1-\text{SYNC}.
\end{equation} 

The discrete nature of the SPIKE-Synchronization profile also requires a different definition for the multivariate case, as discrete functions can not be added in a straightforward manner as done for piecewise constant or piecewise linear functions in \eqref{eqn:avrg_profile}.
For the multivariate profile of $N$ spike-trains, we first define a generalized bivariate coincidence indicator $C^\nind{n,m}_k$ for all pairs of spike trains:
\begin{equation} 
 C^\nind{n,m}_i = \begin{cases} 1\quad\text{if}\quad \min_j(|t^\nind{n}_i - t^\nind{m}_j|) < \tau^\nind{n,m}_{ij}\\
        0\quad\text{otherwise},
       \end{cases}
\end{equation} 
where $\tau^\nind{n,m}_{ij}$ is the same as defined in \eqref{eqn:spike_sync_tau}, but for arbitrary spike trains $n$ and $m$.
From these bivariate coincidence indicators we then compute the average coincidence counter for each spike in every spike train:
\begin{equation} \label{eqn:spike_sync_multivar}
 C^\nind{n}_i = \frac1{N-1}\sum_{m\neq n} C^\nind{n,m}_i.
\end{equation} 
Finally, again all the averaged coincidence counters are merged $\{C^a_k\} = \bigcup_n \{C^\nind{n}_i\}$.
Together with the merged spike times this results in the multivariate SPIKE-Synchronization profile.
Note that in contrast to the bivariate case, the multivariate profile can obtain values different from just zero and one.
Namely, from \eqref{eqn:spike_sync_multivar} we find that $C^a_k = p/(N-1)$ with $p=0\dots N-1$.
Furthermore, note that for $N=2$ the multivariate definition in \eqref{eqn:spike_sync_multivar} becomes equivalent to the bivariate definition of \eqref{eqn:spike_sync_cn}.
Similar to the bivariate case, the overall multivariate SPIKE-Synchronization value is calculated as the ratio of coincident spikes:
\begin{equation}
 \Sync^a = \frac{\sum_k C^a_k}{M^a} = \frac{C^a}{M^a},\qquad D^a_\text{sync} = 1-\text{SYNC}^a,
\end{equation}
where $M^a$ is the overall number of all spikes in the $N$ spike trains.

\subsubsection{Properties}
In contrast to the ISI- and SPIKE-profiles that quantify dissimilarity, the SPIKE-Synchronization profile is a similarity measure.
Furthermore, the SPIKE-Synchronization profiles are defined on discrete points only (the spike times) and are discrete valued as well.
While a bivariate profile can only take two values: zero and one representing the absence or presence of a coincidence, the multivariate profiles can exhibit $N$ values in $[0,1]$. 
It is immediately clear that the SPIKE-Synchronization distance is not a metric, as there exist spike trains with $D_\sync(s_1,s_2)=0$ even for $s_1\neq s_2$.
Furthermore, also transitivity is violated as one easily finds examples where ${D_\sync(s_1,s_2)=0}$ and ${D_\sync(s_2,s_3)=0}$, but ${D_\sync(s_1,s_3)>0}$.

An exemplary multivariate SPIKE-Synchronization profile is shown in \figref{fig:profiles}.
As seen there, SPIKE-Synchronization can clearly differentiate between the noise dominated part at the beginning and the synchronous spikes at the end.
However, at the transition in between, large fluctuations appear due to the discrete nature of the coincidence measure.
Nevertheless, the synchronization events at the beginning are clearly captured as distinguished peaks.

An analytical calculation of the SPIKE-Synchronization distance~$\avrg{D_\sync}$ of Poisson spike trains is performed in Appendix~\ref{app:spike_sync_dist} and gives:
\begin{equation} \label{eqn:D_sync_avrg}
 \avrg{D_\sync} = 1-\frac1{r^{-1}+2+r},
\end{equation}
where $r=\lambda_1/\lambda_2$ is again the ratio of the rates.

\section{Summary and Conclusions} \label{sec:summary}

In this paper, we first provide a comprehensive, concise introduction to the different techniques of spike detection.
This is the basis for any kind of spike train analysis which can help to understand the fundamental processes in the brain.
One approach is to analyze spike train synchrony and for this purpose we here review three time-resolved measures: the ISI- and the SPIKE-distance, which have been known for several years, and the recently proposed SPIKE-Synchronization.
All these methods provide a way to quantify the similarity of spike trains in a local manner, but can also be reduced to an overall distance measure: $D_I$, $D_S$ and $D_\sync$.
Although originally defined only for bivariate profiles of two spike trains, they can be generalized to multivariate situations and measure the combined synchrony of an ensemble of spike trains.

Furthermore, we study the mathematical properties of these three measures in order to provide the necessary information for their successful application to experimental or numerical data.
Specifically, we are able to obtain the expectation values of the overall distances for two Poisson spike trains of arbitrary rates.
This gives another, crucial point of reference for interpreting experimental and numerical results.
For the ISI-distance as well as the SPIKE-Synchronization, we can calculate the expectation values exactly, cf.\ \eqref{eqn:D_isi_poisson} and~\eqref{eqn:D_sync_avrg}.
For the SPIKE-distance, the analytic result is intractable at this point and instead we present an empirical estimate in \eqref{eqn:D_spike_avrg} that shows excellent agreement with numerical results, cf.\ \figref{fig:poisson_distance}.
Note that for all three methods, the expectation value for two Poisson spike trains only depends on the ratio of the rates $r=\lambda_1/\lambda_2$.
This is an immediate consequence of the invariance of the measures under global rescaling of the time.

This detailed mathematical analysis of the spike synchronization measures further improves the usability of these methods for examining and interpreting experimental and numerical results.

We finally note that all three methods are part of the Matlab based graphical user interface SPIKY\footnote{\href{http://www.fi.isc.cnr.it/users/thomas.kreuz/Source-Code/SPIKY.html}{\textsf{http://www.fi.isc.cnr.it/users/thomas.kreuz/Source-Code/SPIKY.html }}}~\cite{kreuz2015SPIKY, bozanic2014SPIKY}.
Furthermore, with the PySpike library\footnote{\href{http://www.pyspike.de}{\textsf{http://www.pyspike.de}}} there also exists a Python implementation for spike train similarity analysis that provides the three methods discussed here.

\section{Acknowledgements}
We thank Daniel Chicharro for preceding calculations on the ISI-Distance and valuable discussions.
We also thank Conor Houghton for useful discussions and comments on the manuscript.
N.B.\ thanks Joshua~D.~Berke for motivational discussions and for hospitality at the University of Michigan.
A.S.\ thanks Filip Melinscak for fruitful discussions and the Institute for Complex Systems, CNR in Sesto Fiorentino for hospitality.

\new{M.M., N.B., A.S.\ and T.K.\ were supported by the European Comission
through the Marie Sklodowska-Curie Initial Training Network ``Neural
Engineering Transformative Technologies (NETT)'' project 289146 and T.K.\ also through the Marie Sklodowska-Curie
European Joint Doctorate ``Complex Oscillatory
Systems: Modeling and Analysis (COSMOS)'' project 642563.}

\appendix
{
\let\normalsize\small
\small

\subsection{ISI-Distance for Poisson Spike Trains} \label{app:isi_dist_poisson}

We are interested in the average ISI-Distance $\avrg{D_I}$ of two Poisson spike trains with rates $\lambda_1$ and $\lambda_2$.
We start by noting that the average ISI-distance is identical to the average profile: $\avrg{D_I} = \avrg{I(t)}$.
Furthermore, for Poisson spike trains with rate $\lambda$ the probability density for the interspike interval $\nu$ is $P(\nu) = \lambda\e^{-\lambda \nu}$.
However, if we consider the interspike intervals as a piecewise constant function~$\nu(t)$, cf.~\eqref{eqn:interspike_intervals}, the probability to be in an interval of length~$\nu(t)$ at time~$t$ is the probability of the interspike interval multiplied by its length:
\begin{equation}
 \tilde P(\nu(t)) = \lambda^2 \nu(t) \e^{-\lambda \nu(t)},
\end{equation} 
where the extra factor $\lambda$ ensures normalization.
For the average profile value we consider the interspike intervals as independent random variables $\nu_{1,2}$ and perform the integration over the probability density:
\begin{align*}
 \avrg{I} &= \int_0^\infty \rmd \nu_1 \int_0^\infty \rmd \nu_2\, I(\nu_1,\nu_2) \tilde P(\nu_1) \tilde P(\nu_2)\\
  &= \int_0^\infty \rmd \nu_1 \int_0^\infty \rmd \nu_2\, \frac{|\nu_1-\nu_2|}{\max\{\nu_1,\nu_2\}} \lambda_1^2 \nu_1 \e^{-\lambda_1 \nu_1} \lambda_2^2 \nu_2 \e^{-\lambda_2 \nu_2}
\end{align*}
The absolute value and the maximum can be resolved by considering $\nu_1>\nu_2$ and $\nu_1<\nu_2$ separately and splitting the integral:
\begin{align*}
 \avrg{I} = &\underbrace{\lambda_1^2 \lambda_2^2 \int_0^\infty \rmd \nu_1 \int_0^{\nu_1} \rmd \nu_2\, (\nu_1-\nu_2) \nu_2 \e^{-\lambda_1 \nu_1 - \lambda_2 \nu_2}}_{\mathcal{I}_1} + \\
  &\underbrace{\lambda_1^2 \lambda_2^2 \int_0^\infty \rmd \nu_1 \int_{\nu_1}^\infty \rmd \nu_2\, (\nu_2-\nu_1) \nu_1 \e^{-\lambda_1 \nu_1 - \lambda_2 \nu_2}}_{\mathcal{I}_2}.
\end{align*}
The first integral evaluates to:
\begin{equation}
 \mathcal{I}_1 = 1 - 2\frac{\lambda_1}{\lambda_2} + 2\frac{\lambda_1^2}{\lambda_2(\lambda_1+\lambda_2)} + \frac{\lambda_1}{(\lambda_1+\lambda_2)^2},
\end{equation}
which can be expressed in terms of the rate ratio $r=\lambda_1/\lambda_2$:
\begin{equation}
 \mathcal{I}_1 = 1-2r+\frac{2r^2}{1+r}+\frac{r^2}{(1+r)^2} = \frac{1}{(1+r)^2}.
\end{equation}
For the second integral, after a change of the order of integration we obtain the same result as above but with $\nu_1\leftrightarrow\nu_2$ interchanged, hence with the inverse ratio $r^{-1}=\lambda_2/\lambda_1$.
Putting these results together we arrive at the average ISI-distance for two Poisson spike trains with a rate ratio $r=\lambda_1/\lambda_2$ (see \figref{fig:poisson_distance}):
\begin{equation} \label{eqn:spike_sync_dist_poisson}
 \avrg{D_I} = \avrg{I(t)} = \mathcal{I}_1 + \mathcal{I}_2 = \frac{1}{(1+r)^2} + \frac{1}{(1+r^{-1})^2} .
\end{equation}

\subsection{SPIKE-Synchronization Distance for Poisson Spike Trains}
\label{app:spike_sync_dist}
Here, we calculate the expectation value of the SPIKE-Synchronization $\avrg{\Sync}$ for two Poisson spike trains with rates $\lambda_1$ and $\lambda_2$.
We start from the coincidence counter $C^\nind{1}_i$ of the first spike train \eqref{eqn:spike_sync_cn}.\footnote{Choosing the first spike train is arbitrary, equivalently the second spike train could be analyzed with the same result.}
Note that it is sufficient to consider only $C^\nind{1}_i$, as there we already account for all possible coincidences.
The overall SPIKE-Synchronization can be calculated as two times the sum of this coincidence counter.
For the average one thus finds:
\begin{equation} \label{eqn:app_e_j}
 \avrg{\Sync} = \avrg{\frac2{M_1+M_2}\sum_{i=1}^{M_1} C^\nind{1}_i} = \frac{2\lambda_1}{\lambda_1+\lambda_2}\avrg{C^\nind{1}_i},
\end{equation}
where $M_{1,2}$ are the number of spikes in the spike trains and in the last step we used that $\avrg{M_{1,2}}=\lambda_{1,2}$ for some spike train length~$T$.

To compute the remaining average, we express $C^\nind{1}_i$ in terms of the following three independent random variables:
\begin{equation}
 \begin{aligned}
  \tilde \tau &= \min\{\nu^\nind{1}_{i}, \nu^\nind{1}_{i-1}, \nu^\nind{2}_{j-1}\}, & P(\tilde\tau) &= \tilde\lambda\e^{-\tilde\lambda\tilde\tau}\\
  \nu &= \nu^\nind{2}_{j}, & P(\nu) &= \lambda_2^2 \nu \e^{-\lambda_2\nu}\\
  x &= (t^\nind{1}_i-t^\nind{2}_j)/\nu^\nind{2}_j,& P(x) &= 1.
 \end{aligned}
\end{equation}
The crucial point is that $\nu^\nind{2}_j$ is removed from the minimum in $\tilde\tau$ and considered separately.
For Poisson spike trains the above variables can be interpreted as follows: for each spike $t^\nind{1}_i$ we randomly choose the minimum of the three surrounding interspike intervals from an exponential distribution with rate $\tilde\lambda = 2\lambda_1+\lambda_2$.
Furthermore, the interspike interval $\nu$ is chosen at random, where we have to take into account that the probability of finding some interval $\nu$ around time $t^\nind{1}_i$ is proportional to $\nu$.
Thus we get $P(\nu)\sim\nu\e^{-\lambda_2\nu}$.
Finally, $x\in[0,1]$ is chosen uniformly to determine the distance between $t^\nind{1}_i$ and $t^\nind{2}_j$.
The coincidence counter is then expressed in terms of these random variables as:
\begin{equation}
 C^\nind{1}_i = \begin{cases} 1\quad \text{if}\quad x\nu \leq \tilde\tau/2 \\
	\;\quad\text{or}\quad (1-x)\nu \leq \tilde\tau/2 \\
	\;\quad\text{or}\quad \nu \leq \tilde\tau \\
        0\quad\text{otherwise}.
       \end{cases} 
\end{equation} 
The average is obtained from integrating over all random variables $\avrg{C^\nind{1}_i} = \int C^\nind{1}_i \rmd\tilde\tau\rmd\nu\rmd x$.
Substituting $C^\nind{1}_i$, the integration over $x$ can be performed immediately.
The remaining integral yields:
\begin{equation}
\begin{aligned}
 \avrg{C^\nind{1}_i}&=\int_0^\infty\rmd\tilde\tau P(\tilde\tau) \left[ \int_0^{\tilde\tau} \rmd \nu P(\nu) + \int_{\tilde\tau}^\infty \rmd\nu \frac{\tilde\tau}{\nu} P(\nu) \right]\\
 &= \frac{\lambda_2}{2(\lambda_1+\lambda_2)}.
\end{aligned}
\end{equation} 
Consequently, using \eqref{eqn:app_e_j} we find for the average overall SPIKE-Synchronization of two Poisson spike trains:
\begin{equation}
 \avrg{\text{SPIKE}} = \frac{\lambda_1\lambda_2}{(\lambda_1+\lambda_2)^2} = \frac1{r+r^{-1}+2},
\end{equation} 
where again the rate ratio $r=\lambda_1/\lambda_2$ is introduced and the result is symmetric in $r$ and $r^{-1}$.
Finally, the SPIKE-Synchronization distance amounts to:
\begin{equation}
 \avrg{D_\sync} = 1-\avrg{\Sync} = 1 - \frac1{r+r^{-1}+2}.
\end{equation} 
This result is numerically validated in \figref{fig:poisson_distance}.
}

%
%
%
\bibliography{ref_EBCCSP15}

\begin{thebibliography}{10}

\bibitem{kandel2012principles}
E.R.~Kandel et~al.
\newblock {\em Principles of Neural Science, 5th edn.}
\newblock The McGraw-Hill Companies, 2012.

\bibitem{quiroga2013principles}
R.~{Quian Quiroga} and S.~Panzeri (eds.).
\newblock {\em Principles of Neural Coding}.
\newblock CRC Taylor and Francis, Boca Raton, FL, 2013.

\bibitem{perkel1968coding}
D.H. Perkel and T.H. Bullock.
\newblock Neural coding.
\newblock {\em Neurosciences Research Program Bulletin}, 3:221--348, 1968.

\bibitem{rieke1996spikes}
F.~Rieke et~al.
\newblock {\em Spikes: Exploring the neural code}.
\newblock Institute of Technology, Cambridge, Massachusetts, 1996.

\bibitem{victor1996nature}
J.D. Victor and K.P. Purpura.
\newblock Nature and precision of temporal coding in visual cortex: a
  metric-space analysis.
\newblock {\em Journal of Neurophysiology}, 76(2):1310--1326, 1996.

\bibitem{rossum2001novel}
M.~{van Rossum}.
\newblock A novel spike distance.
\newblock {\em Neural Computation}, 13(4):751--763, 2001.

\bibitem{quiroga2002event}
R.~{Quian Quiroga} et~al.
\newblock Event synchronization: a simple and fast method to measure
  synchronicity and time delay patterns.
\newblock {\em Physical Review E}, 66(4):041904, 2002.

\bibitem{schreiber2003new}
S.~Schreiber et~al.
\newblock A new correlation-based measure of spike timing reliability.
\newblock {\em Neurocomputing}, 52:925--931, 2003.

\bibitem{schrauwen2007linking}
B.~Schrauwen and J.~{van Campenhout}.
\newblock Linking non-binned spike train kernels to several existing spike
  train metrics.
\newblock {\em Neurocomputing}, 70(7):1247--1253, 2007.

\bibitem{kreuz2007measuring}
T.~Kreuz et~al.
\newblock Measuring spike train synchrony.
\newblock {\em Journal of Neuroscience Methods}, 165(1):151--161, 2007.

\bibitem{kreuz2009measuring}
T.~Kreuz et~al.
\newblock Measuring multiple spike train synchrony.
\newblock {\em Journal of Neuroscience Methods}, 183(2):287--299, 2009.

\bibitem{naud2011improved}
R.~Naud et~al.
\newblock Improved similarity measures for small sets of spike trains.
\newblock {\em Neural Computation}, 23(12):3016--3069, 2011.

\bibitem{kreuz2011time}
T.~Kreuz et~al.
\newblock Time-resolved and time-scale adaptive measures of spike train
  synchrony.
\newblock {\em Journal of Neuroscience Meth.}, 195(1):92--106, 2011.

\bibitem{kreuz2013monitoring}
T.~Kreuz et~al.
\newblock Monitoring spike train synchrony.
\newblock {\em Journal of Neurophysiology}, 109(5):1457--1472, 2013.

\bibitem{rusu2014new}
C.~Rusu and R.~Florian.
\newblock A new class of metrics for spike trains.
\newblock {\em Neural Computation}, 26(2):306--348, 2014.

\bibitem{muino2014frequent}
D.P. Mui{\~n}o and C.~Borgelt.
\newblock Frequent item set mining for sequential data: Synchrony in neuronal
  spike trains.
\newblock {\em Intelligent Data Analysis}, 18(6):997--1012, 2014.

\bibitem{victor2005spike}
J.D. Victor.
\newblock Spike train metrics.
\newblock {\em Current Opinion in Neurobiology}, 15(5):585--592, 2005.

\bibitem{kreuz2015SPIKY}
T.~Kreuz et~al.
\newblock {SPIKY}: A graphical user interface for monitoring spike train
  synchrony.
\newblock {\em Journal of Neurophysiology (accepted)}, 2015.

\bibitem{mcdonough1995detection}
R.N. McDonough.
\newblock {\em Detection of signals in noise}.
\newblock Academic Press, 1995.

\bibitem{gozani1994optimal}
S.N. Gozani and J.~P. Miller.
\newblock Optimal discrimination and classification of neuronal action
  potential waveforms from multiunit, multichannel recordings using
  software-based linear filters.
\newblock {\em Biomedical Engineering, IEEE Transactions on}, 41(4):358--372,
  1994.

\bibitem{guillory1999100}
K.S. Guillory and R.A. Normann.
\newblock A 100-channel system for real time detection and storage of
  extracellular spike waveforms.
\newblock {\em Journal of Neuroscience Methods}, 91(1):21--29, 1999.

\bibitem{chandra1997detection}
R.~Chandra and L.M. Optican.
\newblock Detection, classification, and superposition resolution of action
  potentials in multiunit single-channel recordings by an on-line real-time
  neural network.
\newblock {\em Biomedical Engineering, IEEE Transactions on}, 44(5):403--412,
  1997.

\bibitem{lewicki1998review}
M.S. Lewicki.
\newblock A review of methods for spike sorting: the detection and
  classification of neural action potentials.
\newblock {\em Network: Computation in Neural Systems}, 9(4):R53--R78, 1998.

\bibitem{obeid2004evaluation}
I.~Obeid and P.D. Wolf.
\newblock Evaluation of spike-detection algorithms for a brain-machine
  interface application.
\newblock {\em Biomedical Engineering, IEEE Transactions on}, 51(6):905--911,
  2004.

\bibitem{pouzat2002using}
C.~Pouzat et~al.
\newblock Using noise signature to optimize spike-sorting and to assess
  neuronal classification quality.
\newblock {\em Journal of Neuroscience Methods}, 122(1):43--57, 2002.

\bibitem{quiroga2004unsupervised}
R.~{Quian Quiroga} et~al.
\newblock Unsupervised spike detection and sorting with wavelets and
  superparamagnetic clustering.
\newblock {\em Neural Computation}, 16(8):1661--1687, 2004.

\bibitem{kim2000neural}
K.H. Kim and S.J. Kim.
\newblock Neural spike sorting under nearly 0-db signal-to-noise ratio using
  nonlinear energy operator and artificial neural-network classifier.
\newblock {\em Biomedical Engineering, IEEE Transactions on},
  47(10):1406--1411, 2000.

\bibitem{kaiser1990simple}
J.F. Kaiser.
\newblock On a simple algorithm to calculate the energy of a signal.
\newblock Acoustics, Speech, and Signal Processing, 1990. ICASSP-90., 1990
  International Conference on, 1990.

\bibitem{quiroga2007spike}
R.~{Quian Quiroga}.
\newblock Spike sorting.
\newblock {\em Scholarpedia}, 2(12):3583, 2007.

\bibitem{hulata2002method}
E.~Hulata et~al.
\newblock A method for spike sorting and detection based on wavelet packets and
  shannon's mutual information.
\newblock {\em Journal of Neuroscience Methods}, 117(1):1--12, 2002.

\bibitem{pavlov2007sorting}
A.~Pavlov et~al.
\newblock Sorting of neural spikes: when wavelet based methods outperform
  principal component analysis.
\newblock {\em Natural Computing}, 6(3):269--281, 2007.

\bibitem{harris2000accuracy}
K.D.~Harris et~al.
\newblock Accuracy of tetrode spike separation as determined by simultaneous
  intracellular and extracellular measurements.
\newblock {\em Journal of Neurophysiology}, 84(1):401--414, 2000.

\bibitem{vogelstein2004spike}
R.J.~Vogelstein et~al.
\newblock Spike sorting with support vector machines.
\newblock In {\em Engineering in Medicine and Biology Society, 2004. IEMBS'04.
  26th Annual Int.\ Conf.}, volume~1, pages 546--549. IEEE, 2004.

\bibitem{chicharro2011what}
D.~Chicharro et~al.
\newblock What can spike train distances tell us about the neural code?
\newblock {\em Journal of Neuroscience Methods}, 199:146--165, 2011.

\bibitem{lyttle2011new}
D.~Lyttle and J.M. Fellous.
\newblock A new similarity measure for spike trains: Sensitivity to bursts and
  periods of inhibition.
\newblock {\em Journal of Neuroscience Methods}, 199:296, 2011.

\bibitem{kistler1997reduction}
W.~Kistler et~al.
\newblock {Reduction of the Hodgkin-Huxley equations to a single-variable
  threshold model}.
\newblock {\em Neural Computation}, 9(5):1015--1045, 1997.

\bibitem{bozanic2014SPIKY}
N.~Bozanic et~al.
\newblock {SPIKY}.
\newblock {\em Scholarpedia}, 9(12):32344, 2014.

\end{thebibliography}

\end{document}